\documentclass[a4paper,11pt]{article}
\usepackage{pos}
\usepackage{graphicx} 

\title{Compatibility between $e^+e^-$ and $\tau$ decay data in the di-pion channel and implications for $a_\mu^\mathrm{SM}$ and CVC tests}

\author*[a]{Alejandro Miranda}

\affiliation[a]{Institut de Física d’Altes Energies (IFAE) and The Barcelona Institute of Science and
Technology (BIST), Campus UAB, 08193 Bellaterra (Barcelona), Spain.}

\emailAdd{jmiranda@ifae.es}

\abstract{We have revisited the isospin-breaking corrections relating $\sigma(e^+e^-\to\pi^+\pi^-)$ and $\Gamma(\tau^-\to\pi^-\pi^0\nu_\tau)$~\cite{Castro:2024prg}. We confirm that the associated uncertainty is under control, so that tau data can also be used to predict accurately the leading hadronic contribution to the muon anomalous magnetic moment and precision conserved vector current tests can be carried out.}

\FullConference{10th International Conference on Quarks and Nuclear Physics (QNP2024)\\
8-12 July, 2024\\
Barcelona, Spain\\}

\date{}
\begin{document}

\maketitle

\section{Introduction: }
The measurement of the muon anomalous magnetic moment, $a_\mu=(g_\mu-2)/2$ has achieved an extraordinary precision, thanks to the (perfectly compatible) BNL \cite{Muong-2:2006rrc} and FNAL \cite{Muong-2:2021ojo,Muong-2:2023cdq} results, leading to the combination
\begin{equation}\label{amuexp}
a_\mu^{\mathrm{exp}}=0.00116592059 \pm0.00000000022\,.
\end{equation}
On the contrary, the corresponding Standard Model (SM) prediction is not clear at the moment, due to the uncertainties on the hadronic vacuum polarization piece and, specifically, on its $2\pi$ part. For the Muon g-2 Theory Initiative White Paper \cite{Aoyama:2020ynm}, this contribution was dominated by the combination of the KLOE \cite{KLOE-2:2017fda} and BaBar \cite{BaBar:2012bdw} results, which were in tension (but still at the limit of compatibility, so they could be combined). The White Paper $a_\mu$ prediction is $5.1\,\sigma$ smaller than the experimental average, Eq.~(\ref{amuexp}). Although it was not used for the White Paper, the BMW lattice collaboration obtained a result for $a_\mu^\mathrm{HVP}$ \cite{Borsanyi:2020mff} that is only $1.1\,\sigma$ smaller than $a_\mu^{\mathrm{exp}}$. Its recent improvement combining the lattice simulation with data \cite{Boccaletti:2024guq} achieved accord within $0.9\,\sigma$ with $a_\mu^{\mathrm{exp}}$. In 2023 the situation with $e^+e^-$ data became puzzling, because the new CMD-3 measurement \cite{CMD-3:2023alj,CMD-3:2023rfe} disagrees so much with KLOE, that they cannot be combined, a discrepancy which is not yet understood.

Given this conundrum, we recall that we have advocated since \cite{Miranda:2020wdg} (and emphasized and updated in \cite{Masjuan:2023qsp, Castro:2024prg}) that two-pion tau decay data should be used again~\footnote{The usefulness of $\tau$ data for this purpose was put forward by Alemany, Davier and H\"ocker in Ref.~\cite{Alemany:1997tn}, see also e.g. Refs.~\cite{Cirigliano:2001er,Cirigliano:2002pv, Davier:2002dy, Davier:2003pw, Davier:2010fmf, Davier:2010nc,Davier:2013sfa,Davier:2023fpl}.} to obtain the corresponding contribution to $a_\mu^\mathrm{HVP}$, given the fact that:
\begin{enumerate}
    \item All measurements, by the ALEPH \cite{ALEPH:2005qgp}, Belle \cite{Belle:2008xpe}, CLEO \cite{CLEO:1999dln} and OPAL  \cite{OPAL:1998rrm} collaborations are consistent, within errors.
    \item The uncertainty related to the isospin breaking (IB) corrections required is small enough to make this procedure competitive.
\end{enumerate}
Here we revisit the IB corrections relating $e^+e^-$ and $\tau$ data, with a particular focus on those arising from the different neutral (electromagnetic) and charged (weak) form factors.
\section{IB corrections}
In the data-driven approach, at lowest order (LO), the HVP contribution to  $a_{\mu}$ is~\cite{Gourdin:1969dm}  
\begin{equation}\label{eq.amu_dispersive}
    a_\mu^{\text{HVP, LO}}=\frac{1}{4\pi^3}\int_{m_{\pi^0}}^{\infty}ds \,K(s)\,\sigma_{e^+e^-\to\text{hadrons}(+\gamma)}^{0}(s),
\end{equation}

\noindent
where $\sigma^{0}_{e^+e^-\to \text{hadrons}(+\gamma)}(s)$ is the bare hadronic cross-section at hadrons invariant mass $\sqrt{s}$ with vacuum polarization (VP) effects removed~\cite{Eidelman:1995ny} and $K(s)$ is a smooth kernel enhancing the low-energy effects~\cite{Brodsky:1967sr}.

Alternatively, it can also be obtained from~\footnote{CVC stands for conserved vector current, which relates the $e^+e^-\to\pi^+\pi^-$ and $\tau^-\to\pi^-\pi^0\nu_\tau$ in absence of IB corrections.}
\begin{equation}\label{eq:pipi_cross_section}
\left.\sigma^0_{\pi\pi(\gamma)}\right\vert_\text{CVC}=\left[\frac{K_\sigma(s)}{K_\Gamma(s)}\frac{d\Gamma_{\pi\pi[\gamma]}}{ds}\right]\times\frac{R_{\text{IB}}(s)}{S_{\text{EW}}},
\end{equation}
where the ratio of $K$ functions depends on $G_F$, $V_{ud}$, $\alpha$ and $m_\tau$, and the measured spectrum of the di-pion tau decays~\footnote{New physics could affect the tau decays and not the $e^+e^-$ cross-section into two pions~\cite{Miranda:2018cpf,Cirigliano:2018dyk,Gonzalez-Solis:2020jlh,Cirigliano:2021yto}, nevertheless.}, $\frac{d\Gamma_{\pi\pi[\gamma]}}{ds}$, needs to be corrected for the short-distance universal electroweak radiative corrections, $S_\mathrm{EW}$~\cite{Marciano:1988vm}, and the isospin-breaking (IB) factor
\begin{equation}\label{RIB}
    R_{\text{IB}}(s)=\frac{\text{FSR}(s)}{G_{\text{EM}}(s)}\frac{\beta^3_{\pi^{+}\pi^{-}}(s)}{\beta^3_{\pi^+\pi^0}(s)}\left\vert\frac{F_V(s)}{f_{+}(s)}\right\vert^2.
\end{equation}
$R_{\text{IB}}(s)$ depends on two factors which are straightforward: final-state radiative corrections  (FSR) \cite{Drees:1990te} and the kinematical factor depending on the ratio of $\beta$ functions. However, it also depends on two corrections that are challenging. The first of these corresponds to the long-distance QED virtual plus real photon corrections, encoded in the $G_{\text{EM}}$ function, which has been the focus of Refs.~\cite{Cirigliano:2001er, Cirigliano:2002pv,Flores-Baez:2006yiq,Flores-Tlalpa:2006snz,Miranda:2020wdg}~\footnote{We stick here to the `$\mathcal{O}(p^4)$' result of Ref.~\cite{Miranda:2020wdg}, that uses Resonance Chiral Theory \cite{Ecker:1988te,Ecker:1989yg}, which has been successfully employed in computing different contributions to $a_\mu$ \cite{Cirigliano:2002pv,Roig:2014uja, Guevara:2018rhj, Roig:2019reh,Qin:2020udp,Wang:2023njt,Masjuan:2023qsp, Qin:2024ulb,Estrada:2024cfy}. Noteworthy, this result is consistent with the vector meson dominance one employed by the Orsay group since \cite{Davier:2010fmf}.}. Here we concentrate on the second of such corrections, given by the ratio of the neutral $F_V$ and charged $f_+$ pion form factors, where one of the leading IB corrections, the $\rho-\omega$ mixing entering $F_V$, takes place. In this factor, also the mass and (partial) width differences of the neutral and charged $\rho$ mesons play an important role. Particularly, for the IB induced between the neutral and charged $\rho\to\pi\pi(\gamma)$ channels we rely on the results of Ref.~\cite{Flores-Baez:2007vnd}.

In this contribution we will focus on two observables which are sensitive to IB. We will consider the IB corrections to the $\pi\pi$ contribution to $a_\mu^{\text{HVP, LO}}$ obtained using tau data,
\begin{equation}\label{eq:Delta_a_mu}
    \Delta a_\mu^{\text{HVP, LO}}[\pi\pi,\tau]=\frac{1}{4\pi^3}\int_{4m_\pi^2}^{m_\tau^2}ds\, K(s)\left[\frac{K_\sigma(s)}{K_\Gamma(s)}\frac{d\Gamma_{\pi\pi[\gamma]}}{ds}\right]\left(\frac{R_{\text{IB}}(s)}{S_{\text{EW}}}-1\right),
\end{equation}
and also the modification to the di-pion tau decay branching ratio obtained from the $e^+e^-$ measurement,
\begin{equation}\label{eq:ibtpbr}
     \Delta\mathcal{B}_{\pi\pi^0}^{\text{CVC}}=\mathcal{B}_e\int_{4m_\pi^2}^{m_\tau^2}ds\,\sigma_{\pi^+\pi^-(\gamma)}(s)\mathcal{N}(s)\left(\frac{S_\text{EW}}{R_{\text{IB}}(s)}-1\right),
\end{equation}
where $\mathcal{B}_e$ is the electronic tau decay branching ratio and $\mathcal{N}(s)$ depends on $\alpha$, $V_{ud}$ and $m_\tau$.

\section{Form factor parametrizations}
We considered different descriptions of the electromagnetic, $F_V$, and weak, $f_+$, pion form factors (see Ref.~\cite{Castro:2024prg} for details). As typically done by the experimental collaborations, we employed the Gounaris-Sakurai (GS) \cite{Gounaris:1968mw} and K\"uhn-Santamar\'ia (KS) \cite{Kuhn:1990ad} form factors. We also used the Guerrero-Pich (GP) \cite{Guerrero:1997ku} parametrization. However, since it is limited to the $\rho(770)$ resonance, we considered its extension including the $\rho'$ and $\rho''$ resonances, along the lines of Refs.~\cite{GomezDumm:2013sib,Gonzalez-Solis:2019iod}, to construct the input phase shift~\footnote{We also used the form factor giving this phase, which we labelled `Seed'.} of an unsubtracted dispersive form factor, where we accounted for $\rho-\omega-\phi$ mixing (we verified that complex mixing coefficients were needed) and inelastic effects (captured by a conformal polynomial), following Ref.~\cite{Colangelo:2018mtw}.

\section{Results}
We have computed the IB corrections in Eqs.~(\ref{eq:Delta_a_mu}) and (\ref{eq:ibtpbr}), according to the different form factor parametrizations: GS, KS, GP, Seed and Dispersive. Both the fits to data and the analyticity tests (see Ref.~\cite{Castro:2024prg}) work best (and better when KLOE is excluded) for GS and Dispersive. Taking this into account, we will take the latter as our reference result and add linearly a systematic uncertainty coming from its difference with GS in our final results.

In Table \ref{tab:IB_BR_CMD3} we collect the different IB corrections to the CVC di-pion tau decay branching ratio, obtained using the different form factors considered, and we split them according to their type. Focusing on GS and Dispersive, there is good agreement in all contributions (but for the individual effect of the $\rho$ mass difference and the $\rho-\omega$ interference, which are compensated in their sum, however) and the final results agree nicely.

\begin{table}[ht]
    \centering
    \resizebox{0.90\textwidth}{!}{\begin{tabular}{cccccc}
    \hline
        Source & \multicolumn{5}{c}{$\Delta \mathcal{B}_{\pi\pi}^{\text{CVC}}\,(10^{-2})$}\\[0.3ex]
          & GS & KS & GP & Seed & Dispersive \\[0.3ex]
          &  &  &  &  & $p_{4-1} $\\[0.3ex]
        \hline
        $S_\text{EW}$ & \multicolumn{5}{c}{$+0.57(1)$} \\[0.3ex]
        $G_\text{EM}$ &  \multicolumn{5}{c}{$-0.09(^{3}_{1})$} \\[0.3ex]
        FSR & \multicolumn{5}{c}{$-0.19(2)$} \\[0.3ex]
        $m_{\pi^\pm}-m_{\pi^0}$ effect on $\sigma$ & \multicolumn{5}{c}{$+0.20$} \\[0.3ex]
        $m_{\pi^\pm}-m_{\pi^0}$ effect on $\Gamma_\rho$ & $-0.21$ & $-0.22$ & $-0.22$ & $-0.23$ & $-0.20$ \\[0.3ex]
        $m_{K^\pm}-m_{K^0}$ effect on $\Gamma_\rho$ & $-$ & $-$ & $-0.02$ & $-0.03$ &  $+0.01$\\[0.3ex]
        $m_{\rho^\pm}-m_{\rho^0}$ & $+0.08(8)$ & $+0.09(8)$ & $-0.02(2)$ & $-0.02(2)$ & $-0.02(^{2}_{1})$ \\[0.3ex]
        $\rho-\omega$ interference & $-0.08(0)(^{12}_{\,\,0})$ & $-0.09(0)(^{11}_{\,\,0})$ & $-0.09(0)(^{16}_{\,\,1})$ & $-0.06(0)(^{10}_{\,\,0})$ & $-0.01(0)(^{6}_{4})$ \\[0.3ex]
        $\rho-\phi$ interference & $-0.00(0)(_{0}^{1})$ & $-0.00(0)(0)$ & $-$ & $-0.01(0)(0)$ & $-0.01(0)(1)$\\[0.3ex]
        $\pi\pi\gamma$, electromagnetic decays & $+0.34(3)$ & $+0.37(4)$ & $+0.34(4)$ & $+0.34(4)$ & $+0.37(4)$ \\[0.3ex]
        \hline
        TOTAL & $+0.62(^{15}_{\,\,9})$ & $+0.64(^{15}_{\,\,9})$ & $+0.48(^{17}_{\,\,5})$ & $+0.48(^{12}_{\,\,5})$ & $+0.63(^{8}_{6})(^{0}_{3})$ \\[0.3ex]
        \hline
    \end{tabular}}
    \captionsetup{width=0.88\linewidth}
    \caption{IB contributions to BR($\tau^-\to\pi^-\pi^0\nu_\tau$) according to the different form factor inputs. The uncertainties are mostly of systematic origin and are specified in Ref.~\cite{Castro:2024prg}. In the last entry, we take as an additional uncertainty (last shown) the difference between our preferred option for the conformal polynomial $(p_{4-1})$ and the other dispersive results that we considered~\cite{Castro:2024prg}.}
    \label{tab:IB_BR_CMD3}
\end{table}

In Table \ref{tab:IB_amu_CMD3} we display the different IB corrections to the di-pion contribution to $a_\mu^{\text{HVP, LO}}$ using the diverse FF considered. Concentrating on the GS and dispersive results, we made analogous observations as with Table \ref{tab:IB_BR_CMD3}, highlighting again the good consistency of our reference results.

\begin{table}[ht]
    \centering
    \resizebox{0.90\textwidth}{!}{\begin{tabular}{cccccc}
    \hline
    Source & \multicolumn{5}{c}{$\Delta a_\mu^\text{had,LO}[\pi\pi,\tau]\,(10^{-10})$} \\[0.3ex]
     & GS & KS & GP & Seed & Dispersive\\[0.3ex]
      &  &  &  &  & $p_{4-1}$ \\[0.3ex]
    \hline
    $S_\text{EW}$ & \multicolumn{5}{c}{$-11.96(0.15)$} \\[0.3ex]
    $G_\text{EM}$ & \multicolumn{5}{c}{$-1.71(^{0.61}_{1.48})$} \\[0.3ex]
    FSR & \multicolumn{5}{c}{$+4.56(0.46)$} \\[0.3ex]
    $m_{\pi^\pm}-m_{\pi^0}$ effect on $\sigma$ & \multicolumn{5}{c}{$-7.47$} \\[0.3ex]
    $m_{\pi^\pm}-m_{\pi^0}$ effect on $\Gamma$ & $+3.74$ & $+4.12$ & $+4.07$ & $+4.13$ & $+3.58$ \\[0.3ex]
    $m_{K^\pm}-m_{K^0}$ effect on $\Gamma$ & $-$ & $-$ & $+0.37$ & $+0.36$ & $-0.22$ \\[0.3ex]
    $m_{\rho^\pm}-m_{\rho^0}$ & $+0.10(^{0.18}_{0.09})$ & $-0.04(^{0.06}_{0.00})$ & $+1.87(^{1.75}_{1.68})$ & $+1.86(^{1.75}_{1.68})$ & $+1.85(^{1.69}_{1.66})$ \\[0.3ex]
    $\rho-\omega$ interference & $+3.84(0.08)(^{0.23}_{2.17})$ & $+4.00(0.08)(_{1.96}^{0.25})$ & $+4.33(0.07)(^{0.33}_{2.83})$ & $+3.57(0.07)(^{0.18}_{1.77})$ & $+2.72(0.08)(^{0.65}_{1.15})$ \\[0.3ex]
    $\rho-\phi$ interference & $+0.09(0.03)(^{0.06}_{0.10})$ & $+0.03(0.03)(^{0.03}_{0.05})$ & $-$ & $+0.13(0.03)(^{0.05}_{0.07})$ & $+0.12(0.02)(^{0.14}_{0.08})$ \\[0.3ex]
    $\pi\pi\gamma$ & $-6.09(0.67)$ & $-6.68(0.74)$ & $-6.19(0.68)$ & $-6.21(0.68)$ & $-6.62(0.73)$  \\[0.3ex]
    \hline
    TOTAL & $-14.90(^{1.08}_{2.76})$ & $-15.15(^{1.11}_{2.61})$ & $-12.13(^{2.06}_{3.70})$ & $-12.74(^{2.04}_{2.97})$ & $-15.15(^{2.11}_{2.65})(^{0.24}_{0.09})$ \\[0.3ex]
    \hline
    \end{tabular}}
    \captionsetup{width=\linewidth}
    \caption{Contributions to $a_\mu^{\text{had,LO}}[\pi\pi,\tau](10^{-10})$ from the isospin-breaking (IB) corrections according to the different form factor inputs. The uncertainties are mostly systematic, and are discussed in Ref.~\cite{Castro:2024prg}. In the last entry, we take as an additional uncertainty (last shown) the difference between $p_{4-1}$ and the other dispersive results that we considered.} 
    \label{tab:IB_amu_CMD3}
\end{table}

\begin{figure}[ht]
    \centering
    \includegraphics[width=0.48\textwidth]{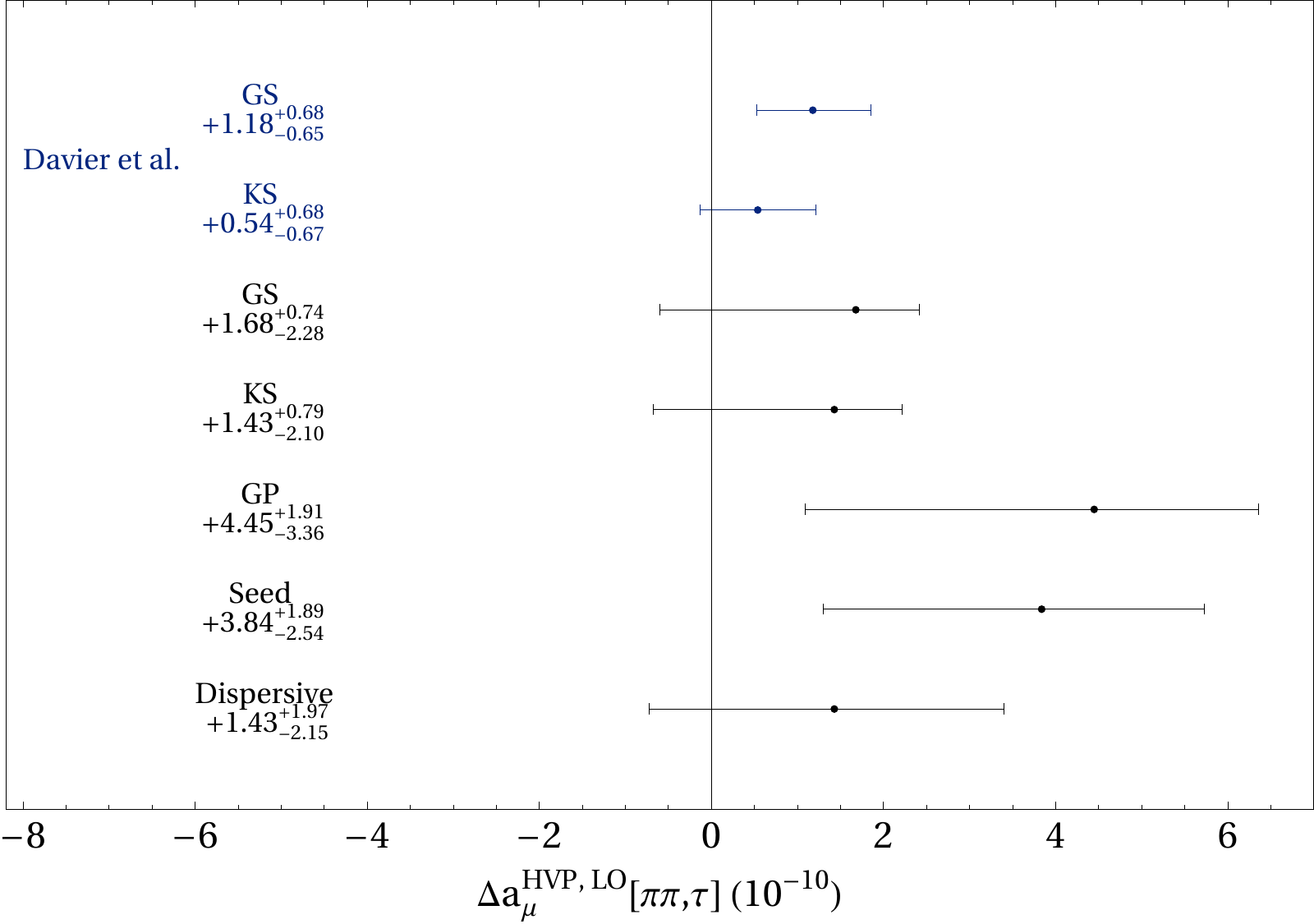}
    \includegraphics[width=0.494\textwidth]{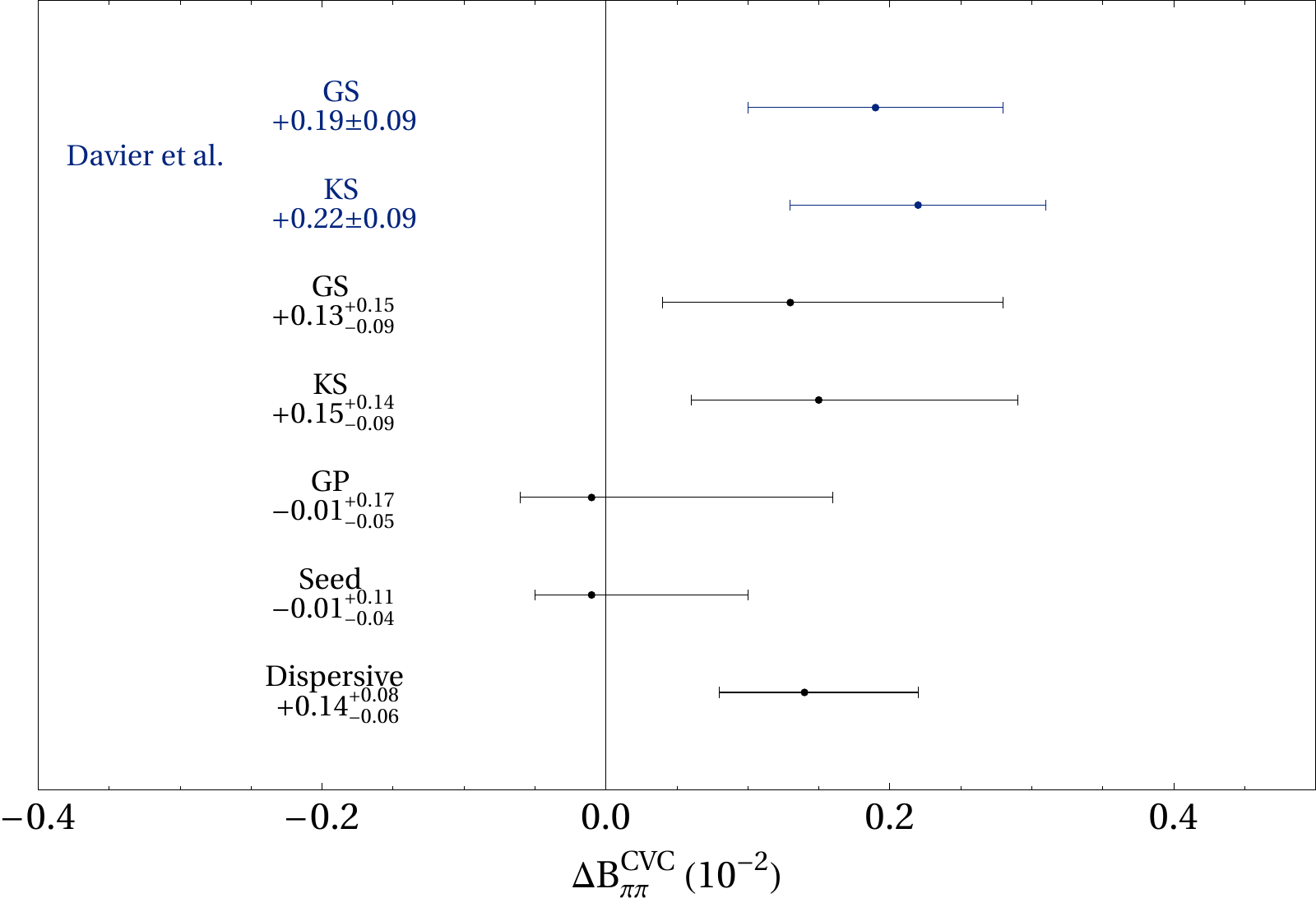}
    \caption{IB corrections in the ratio of the form factors $\vert F_V(s)/f_+(s)\vert$ to $a_\mu^\text{HVP, LO}$ and $\mathcal{B}^\text{CVC}_{\pi\pi}$.}
    \label{fig:IB_correctionsFF}
\end{figure}

In Fig.~\ref{fig:IB_correctionsFF} we show the IB corrections to the di-pion tau decay branching ratio and contribution to $a_\mu^{\text{HVP, LO}}$, using the inputs in Tables \ref{tab:IB_BR_CMD3} and \ref{tab:IB_amu_CMD3}, where we also compare to the results in Ref.~\cite{Davier:2010fmf} (in blue).

\begin{figure}[ht]
    \centering 
    \includegraphics[width=0.88\textwidth]{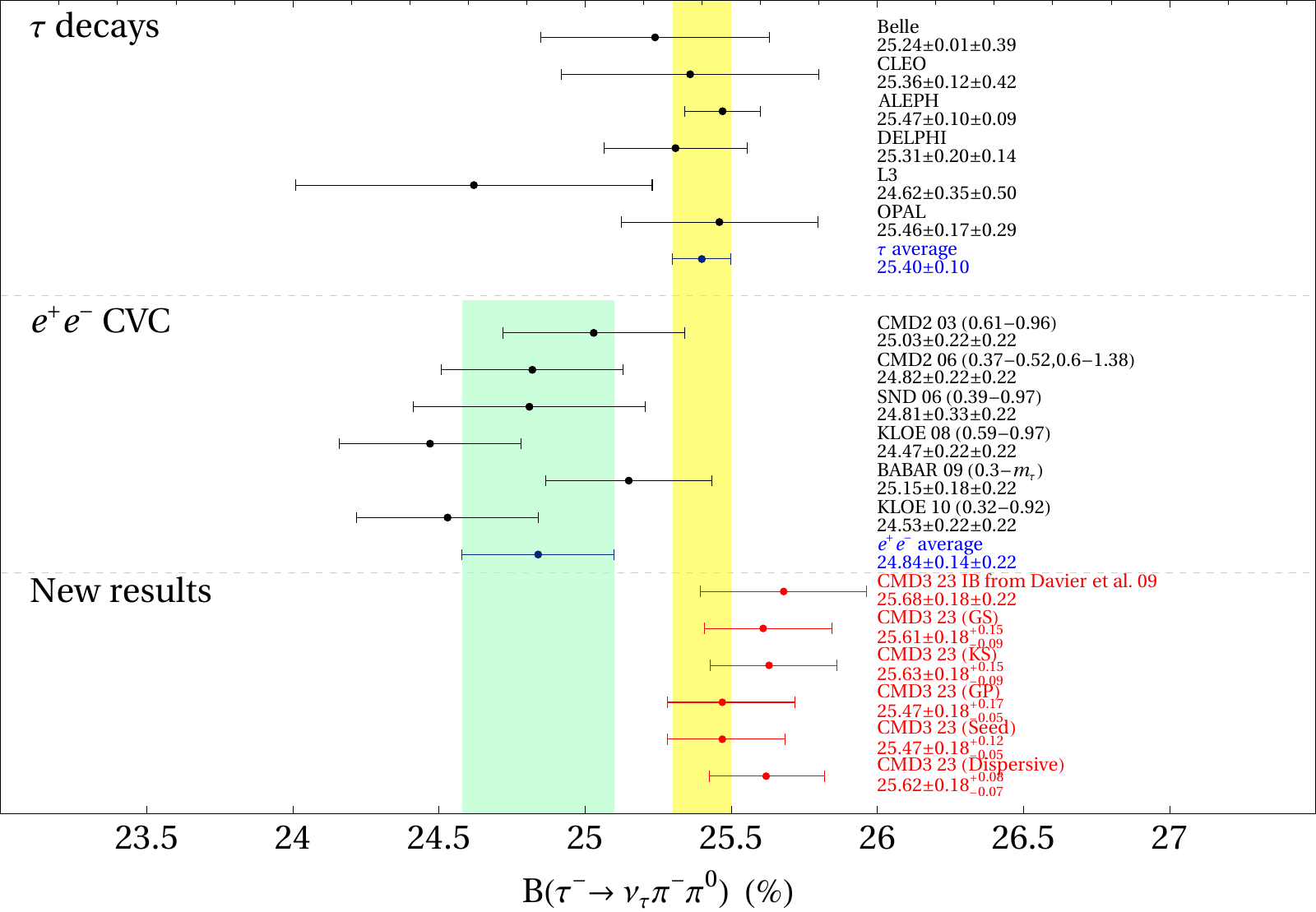}
    \captionsetup{width=0.88\linewidth}
    \caption{Comparison between the measured branching fractions for $\tau^-\to\pi^-\pi^0\nu_\tau$ and the prediction from the $e^+e^-\to\pi^+\pi^-$ spectral functions, applying the isospin-breaking corrections given in Table \ref{tab:IB_BR_CMD3}. Our reference result comes from the dispersive evaluation, and we add linearly to the final theory uncertainty (second error shown, preceded by the statistical one) its difference with the GS value, that also complies well with analyticity.    }
    \label{fig:BR_CVC}
\end{figure}

Finally, in Fig.~\ref{fig:BR_CVC} we compare the measured di-pion tau decay branching fractions and the predictions from the $e^+e^-\to\pi^+\pi^-$ spectral functions, applying the IB corrections given in Table \ref{tab:IB_BR_CMD3}. We remark the very good consistency between all results using the CMD-3 data and updated IB corrections (even more for our preferred results, GS and dispersive, and those of Ref.~\cite{Davier:2023fpl}).

\section{Conclusions}
We have revisited the IB corrections relating the $e^+e^-$ and $\tau$ decay di-pion observables, particularly focusing on the one coming from the ratio of the electromagnetic and weak form factors. We have considered the popular parametrizations Gounaris-Sakurai, K\"uhn-Santamar\'ia, Guerrero-Pich (and its extension with excited isovector mesons) and a dispersive form factor, favoring the first and last one through fits to data and analyticity tests (detailed in Ref.~\cite{Castro:2024prg}). For our main observables of interest, we find $\Delta\mathcal{B}_{\pi\pi}^\mathrm{CVC}=\left(+0.63_{-0.08}^{+0.09}\right)\times10^{-2}\,,\; \Delta a_\mu^{\mathrm{had,LO}[\pi\pi,\tau]}=\left(-15.15^{+2.37}_{-2.90}\right)\times10^{-10}$, corresponding to $\Delta a_\mu=a_\mu^{\mathrm{exp}}-a_\mu^{\mathrm{SM}}=\left(14.8^{+5.1}_{-5.4}\right)\times10^{-10}$, a $2.7\,\sigma$ difference, with agrees nicely with Refs.~\cite{Davier:2010fmf, Davier:2023fpl}. Our analysis confirms the reliability of these IB corrections, supporting the use of tau data in the updated SM prediction of $a_\mu$.

\acknowledgments
We gladly acknowledge the QNP24 organizing committees. We thank financial support through the Ministerio de Ciencia e Innovación under Grant No. PID2020–112965 GB-I00, by the Departament de Recerca i Universitats from Generalitat de Catalunya to the Grup de Recerca ``Grup de Física Teòrica UAB/IFAE'' (Codi: 2021 SGR 00649) and partial Conahcyt support, from project CB2023-2024-3226.

\end{document}